\documentclass[twocolumn]{aastex63}
\usepackage{hyperref}
\usepackage[colorinlistoftodos]{todonotes}
\usepackage{amsmath}
\usepackage{nicefrac}
\shorttitle{Dusty Quiescents}
\usepackage[encapsulated]{CJK}

\newcommand{\MsolPerYr}{$M_\odot~\mathrm{yr}^{-1}$}

\newcommand{\microns}{$\mu$m}
\newcommand{\sersic}{S\'{e}rsic}

\begin{document}

\title{UNCOVER: Significant Reddening in Cosmic Noon Quiescent Galaxies}

\correspondingauthor{Jared Siegel}
\email{siegeljc@princeton.edu}

\author[0000-0002-9337-0902]{Jared C. Siegel}
\altaffiliation{NSF Graduate Research Fellow}
\affiliation{Department of Astrophysical Sciences, Princeton University, 4 Ivy Lane, Princeton, NJ 08544, USA}

\author[0000-0003-4075-7393]{David J. Setton}
\altaffiliation{Brinson Prize Fellow}
\affiliation{Department of Astrophysical Sciences, Princeton University, 4 Ivy Lane, Princeton, NJ 08544, USA}

\author[0000-0002-5612-3427]{Jenny E. Greene}
\affiliation{Department of Astrophysical Sciences, Princeton University, 4 Ivy Lane, Princeton, NJ 08544, USA}

\author[0000-0002-1714-1905]{Katherine A. Suess}
\affiliation{Department for Astrophysical \& Planetary Science, University of Colorado, Boulder, CO 80309, USA}

\author[0000-0001-7160-3632]{Katherine E. Whitaker}
\affiliation{Department of Astronomy, University of Massachusetts, Amherst, MA 01003, USA}
\affiliation{Cosmic Dawn Center (DAWN), Denmark}

\author[0000-0001-5063-8254]{Rachel Bezanson}
\affiliation{Department of Physics and Astronomy and PITT PACC, University of Pittsburgh, Pittsburgh, PA 15260, USA}

\author[0000-0001-6755-1315]{Joel Leja}
\affiliation{Department of Astronomy \& Astrophysics, The Pennsylvania State University, University Park, PA 16802, USA}
\affiliation{Institute for Computational \& Data Sciences, The Pennsylvania State University, University Park, PA 16802, USA}
\affiliation{Institute for Gravitation and the Cosmos, The Pennsylvania State University, University Park, PA 16802, USA}

\author[0000-0001-6278-032X]{Lukas J. Furtak}
\affiliation{Physics Department, Ben-Gurion University of the Negev, P.O. Box 653, Be'er-Sheva 84105, Israel}

\author[0000-0002-7031-2865]{Sam E. Cutler}
\affiliation{Department of Astronomy, University of Massachusetts, Amherst, MA 01003, USA}

\author[0000-0002-2380-9801]{Anna de Graaff}
\affiliation{Max-Planck-Institut f\"ur Astronomie, K\"onigstuhl 17, D-69117, Heidelberg, Germany}

\author[0000-0002-1109-1919]{Robert Feldmann}
\affiliation{Department of Astrophysics, University of Zurich, CH-8057, Switzerland}

\author[0000-0002-3475-7648]{Gourav Khullar}
\affiliation{Department of Physics and Astronomy and PITT PACC, University of Pittsburgh, Pittsburgh, PA 15260, USA}

\author[0000-0002-2057-5376]{Ivo Labbe}
\affiliation{Centre for Astrophysics and Supercomputing, Swinburne University of Technology, Melbourne, VIC 3122, Australia}

\author[0000-0001-9002-3502]{Danilo Marchesini}
\affiliation{Department of Physics \& Astronomy, Tufts University, MA 02155, USA}

\author[0000-0001-8367-6265]{Tim B. Miller}
\affiliation{Center for Interdisciplinary Exploration and Research in Astrophysics (CIERA), Northwestern University,1800 Sherman Ave, Evanston, IL 60201, USA}

\author[0000-0003-2804-0648 ]{Themiya Nanayakkara}
\affiliation{Centre for Astrophysics and Supercomputing, Swinburne University of Technology, Melbourne, VIC 3122, Australia}

\author[0000-0002-9651-5716]{Richard Pan}\affiliation{Department of Physics \& Astronomy, Tufts University, MA 02155, USA}

\author[0000-0002-0108-4176]{Sedona H. Price}
\affiliation{Department of Physics and Astronomy and PITT PACC, University of Pittsburgh, Pittsburgh, PA 15260, USA}

\author[0000-0003-0660-9776]{Helena P. Treiber}
\altaffiliation{NSF Graduate Research Fellow}
\affiliation{Department of Astrophysical Sciences, Princeton University, 4 Ivy Lane, Princeton, NJ 08544, USA}

\author[0000-0002-8282-9888]{Pieter van Dokkum}
\affiliation{Department of Astronomy, Yale University, New Haven, CT 06511, USA}

\author[0000-0001-9269-5046]{Bingjie Wang (\begin{CJK*}{UTF8}{gbsn}王冰洁\ignorespacesafterend\end{CJK*})}
\affiliation{Department of Astronomy \& Astrophysics, The Pennsylvania State University, University Park, PA 16802, USA}
\affiliation{Institute for Computational \& Data Sciences, The Pennsylvania State University, University Park, PA 16802, USA}
\affiliation{Institute for Gravitation and the Cosmos, The Pennsylvania State University, University Park, PA 16802, USA}

\author[0000-0003-1614-196X]{John R. Weaver}
\affiliation{Department of Astronomy, University of Massachusetts, Amherst, MA 01003, USA}

\begin{abstract}

We explore the physical properties of five massive quiescent galaxies at $z\sim2.5$, revealing the presence of non-negligible dust reservoirs.
JWST NIRSpec observations were obtained for each target, finding no significant line emission; 
multiple star formation tracers independently place upper limits between $0.1-10$~{\MsolPerYr}.
Spectral energy distribution modeling with \texttt{Prospector} infers stellar masses between $\log_{10}[M / M_\odot] \sim 10-11$ and stellar mass-weighted ages between $1-2$~Gyr.
The inferred mass-weighted effective radii ($r_\mathrm{eff}\sim 0.4-1.4$~kpc) and inner $1$~kpc stellar surface densities ($\log_{10}[\Sigma_{<1\mathrm{kpc}} / {M_\odot~\mathrm{kpc}^2} ]\gtrsim 9$) are typical of quiescent galaxies at $z \gtrsim 2$.
The galaxies display negative color gradients (redder core and bluer outskirts); for one galaxy, this effect results from a dusty core, while for the others it may be evidence of an ``inside--out" growth process.
Unlike local quiescent galaxies, we  identify significant reddening in these typical cosmic noon passive galaxies; 
all but one require $A_V \gtrsim 0.4$. 
This finding is in qualitative agreement with previous studies but our deep 20--band NIRCam imaging is able to significantly suppress the dust--age degeneracy and confidently determine that these galaxies are reddened.
We speculate about the physical effects that may drive the decline in dust content in quiescent galaxies over cosmic time.

\end{abstract}


\section{Introduction}
\label{sec:intro}

Galaxies broadly fall into two classes: star forming and quiescent.
How and when galaxies cease star formation (i.e., transition from star forming to quiescent) remains an open area of study.
In the standard picture, local ellipticals grow ``inside--out"---i.e., by accretion onto massive quenched cores \citep{Bezanson2009}.
Indeed, archaeology studies have revealed that the cores of local massive quiescent galaxies are old
\citep[$\gtrsim10$~Gyr,][]{Trager2000,Thomas2005,SanchezBlazquez2006,Graves2007, Greene2013, McDermid2015} and compact \citep{Shen2003, Trujillo2004, Cappellari2013}.
Several studies have successfully explained these local measurements by forward modeling $z \gtrsim 1$ quiescent galaxies \citep[e.g.,][]{Wuyts2010,Mosleh2017,Suess2019b, Suess2021}. 
However, local quiescent galaxies only tell part of the story; 
half of massive galaxies are thought to have quenched within the Universe's first $3$~Gyr \citep[$z \gtrsim 2$,][]{Muzzin2013}. 
Mapping how and when galaxies quench therefore requires a census of quiescent galaxies across cosmic time. 

The processes that govern galaxy quenching remain poorly understood. 
One explanation for the cessation of star formation is gas depletion \citep{Dave2012,Lilly2013,Man2018}. 
Indeed, quiescent galaxies are predominantly gas and dust poor in the local Universe \citep{Smith2012,Saintonge2017,Donevski2023};
there are notable exceptions with centrally concentrated dust and/or dust lanes, likely resulting from recent mergers \citep[e.g.,][]{Ebneter1985}. 
Several quenched galaxies at $z\gtrsim1$ are also known to be gas depleted \citep{Sargent2015, Bezanson2019,Williams2021}, and neutral outflows appear quite common in high-z post-starburst galaxies \citep{Belli2024,Park2024,Davies2024}.
However, several alternative quenching channels have also been identified---e.g., stabilization by stellar spheroids \citep{Martig2009, Tacchella2015} and environmental effects \citep{Peng2010}. 
The relative importance of the different quenching channels, as a function of mass, environment, and redshift, remains an open question.
Tracing the ISM content of quenched galaxies across cosmic time is therefore a critical probe of how and when galaxies cease star formation \citep{Gobat2018,Magdis2021,Whitaker2021}.

By measuring dust reddening, stellar population synthesis modeling \citep[e.g.,][]{Conroy2013} offers valuable insights into the ISM content of quenched galaxies. 
Prior observations suggest quiescent galaxies at cosmic noon may undergo significant dust attenuation (unlike local quiescent galaxies).
Ground-based NIR photometry (UltraVISTA and ZFOURGE) hints to the typical reddening of quiescent galaxies reaching $A_V \sim 0.5$ at $z \sim 2$ \citep{Marchesini2014,Straatman2016,Martis2019}.
Spatially resolved HST grism spectroscopy \citep[REQUIEM--2D,][]{Akhshik2020} of $z\sim2$ lensed quiescent galaxies also favors non-zero dust attenuation \citep[typically $A_V\lesssim1$,][]{Akhshik2023}.
Measurements of dust attenuation are notoriously challenging for quiescent galaxies, particularly at higher redshifts.
Reddening, stellar age, and metallicity are largely degenerate in stellar population synthesis modeling \citep{Conroy2013}. 
Suppressing this degeneracy requires photometric coverage significantly redward of the Balmer break (i.e., rest-frame optical and NIR) to rule out maximally old stellar populations.
Measuring the dust attenuation of quiescent galaxies is therefore particularly daunting at $z\gtrsim2$.

With high angular resolution and rest-frame coverage from $0.2-1.4~${\microns} ($z \sim 2.5$), JWST NIRCam is well poised to suppress the dust--age degeneracy.
Already, JWST has detected several reddened quiescent galaxies at $z \gtrsim 2$ \citep{Rodighiero2023,Alberts2023,Nanayakkara2024,Setton2024,Lee2024}. 
In this paper, we consider a sample of five spectroscopically confirmed massive quiescent galaxies at $z \sim 2.5$, with 20--band NIRCam imaging. 
We lift the dust--age degeneracy and conclude the galaxies are indeed quenched and significantly reddened.

We present our observations and methods in Sections~\ref{sec:obs}~and~\ref{sec:methods}.
Our results are reported in Section~\ref{sec:results}.
We discuss our findings in Section~\ref{sec:discussion} and conclude in Section~\ref{sec:future}.
Throughout this paper, we adopt the cosmological parameters from the WMAP~9 year results \citep{Hinshaw2013}: $H_0 = 69.32 \ \mathrm{km \ s^{-1} \ Mpc^{-1}}$, $\Omega_m = 0.2865$, and $\Omega_\Lambda = 0.7135$.
We use the Chabrier initial mass function \citep{Chabrier2003} and AB magnitudes.
Unless otherwise stated, all stellar masses and physical scales are lensing corrected.

\section{Observations}
\label{sec:obs}

We consider JWST imaging and spectroscopy of cluster Abell 2744 from the UNCOVER and MegaScience programs. 
The UNCOVER Treasury Program \citep[PIs: Labb\'e and Bezanson, JWST-GO-2561,][]{Bezanson2022} obtained NIRCam imaging in seven bands (F115W, F150W, F200W, F277W, F356W, F410M, and F444W) of $\gtrsim 40$~arcminutes$^2$ to point-source AB depths of $29-30$~magnitude; 
approximately $700$ sources were targeted for NIRSpec spectroscopy (observed-frame $0.6-5.3$~{\microns}).
The MegaScience program \citep[PI: Suess, JWST-GO-4111,][]{Suess2024} imaged approximately $30$~arcminutes$^2$ of Abell 2744 with two wide bands (F070W and F090W) and eleven medium bands (F140M, F162M, F182M, F210M, F250M, F300M, F335M, F360M, F430M, F460M, and F480M).
The reduction pipelines were presented in \cite{Bezanson2022} and \cite{Suess2024}.
Throughout this work, we use the v7.2 imaging mosaics hosted on the DAWN JWST Archive \citep{Valentino2023}.\footnote{\url{https://dawn-cph.github.io/dja/imaging/v7/}}

\cite{Weaver2024} produced photometric catalogs from  joint UNCOVER and archival HST mosaics.
For target selection, we adopt the UNCOVER DR3 catalog;\footnote{\url{https://jwst-uncover.github.io/DR3.html}} 
catalog photometry was extracted from circular apertures on PSF-matched images (aperture sizes were scaled to the galaxies' sizes in the segmentation map). 
Rest-frame colors, stellar masses, and star formation rates are available from \texttt{Prospector}--$\beta$ fits for all sources in the catalog \citep{Wang2023,Wang2024}.
Throughout this work, the catalog photometry and their corresponding stellar population models are only used for target selection; our custom photometric extractions and stellar population synthesis fits are presented in  Section~\ref{sec:methods}.
Gravitational lensing by Abell 2744 is inferred and corrected following the analytic model by \cite{Furtak2023}, recently updated with new spectroscopic redshifts \citep{Price2024}.

\subsection{Sample Selection}
\label{sec:sample}

\begin{figure}[t!]
\gridline{\fig{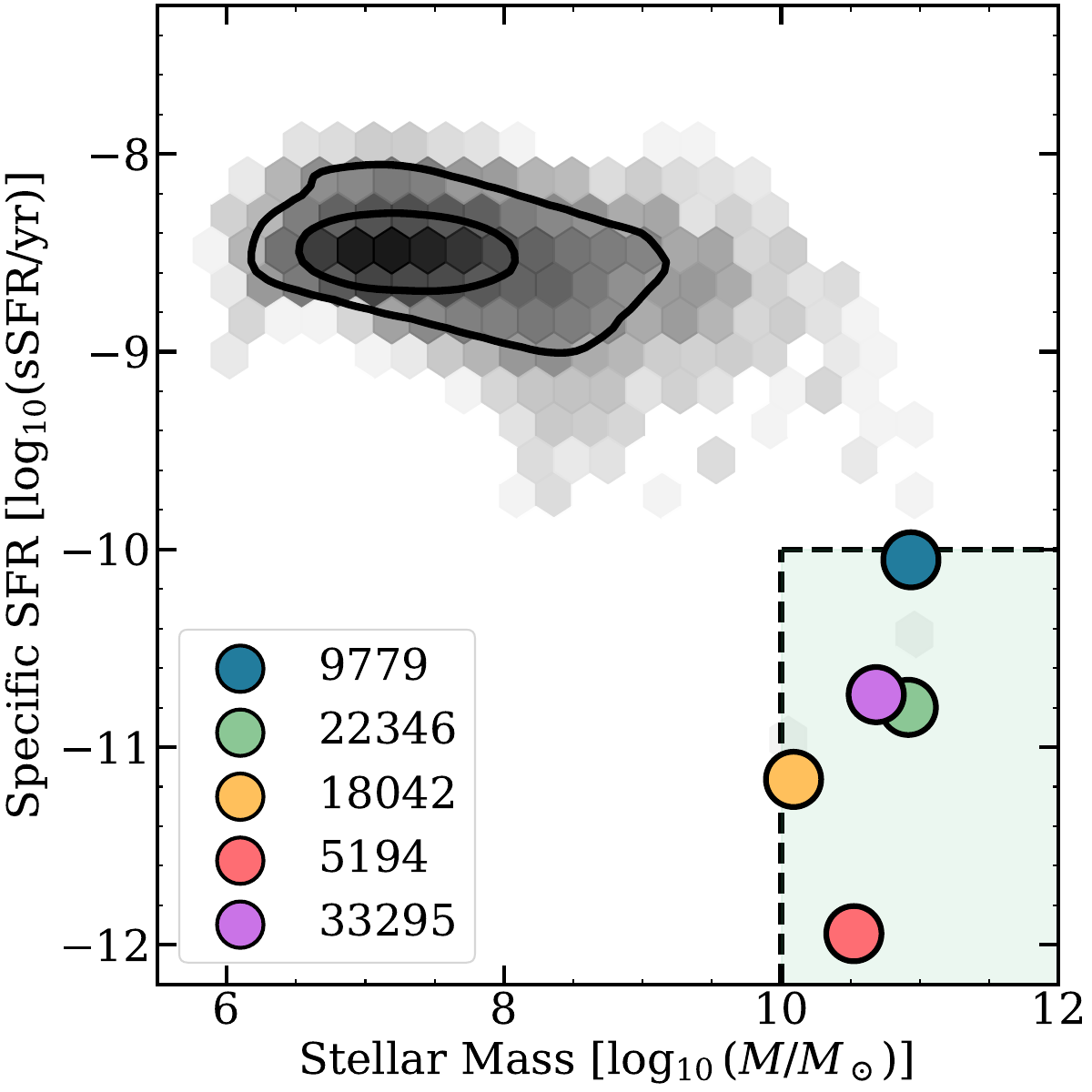}{
\columnwidth}{ } }
\caption{
We select all spectroscopically confirmed cosmic noon massive quiescent galaxies from the UNCOVER survey.
In particular, we select spectroscopically confirmed galaxies satisfying (i) $\log_{10}[ \frac{\mathrm{sSFR}}{\mathrm{yr}}] < -10$, (ii) $\log_{10}[ \frac{M}{M_\odot}] > 10$, and (iii) $2 < z < 3$, based on the publicly available \texttt{Prospector}--$\beta$ fits to the UNCOVER DR3 catalog photometry.
The specific star formation rates and stellar masses for our selected sample are presented, alongside the distribution of $2 < z < 3$ galaxies in the UNCOVER DR3 catalog;
the parent distribution is shown as a 2--dimensional histogram (logarithmic shading) with the $50$ and $84$th percentiles presented as solid lines.
Our independent stellar mass and specific star formation rate measurements are presented in Table~\ref{tab:results}.
}
\label{fig:collective_selection}
\end{figure}

\begin{figure*}[t!]
\gridline{\fig{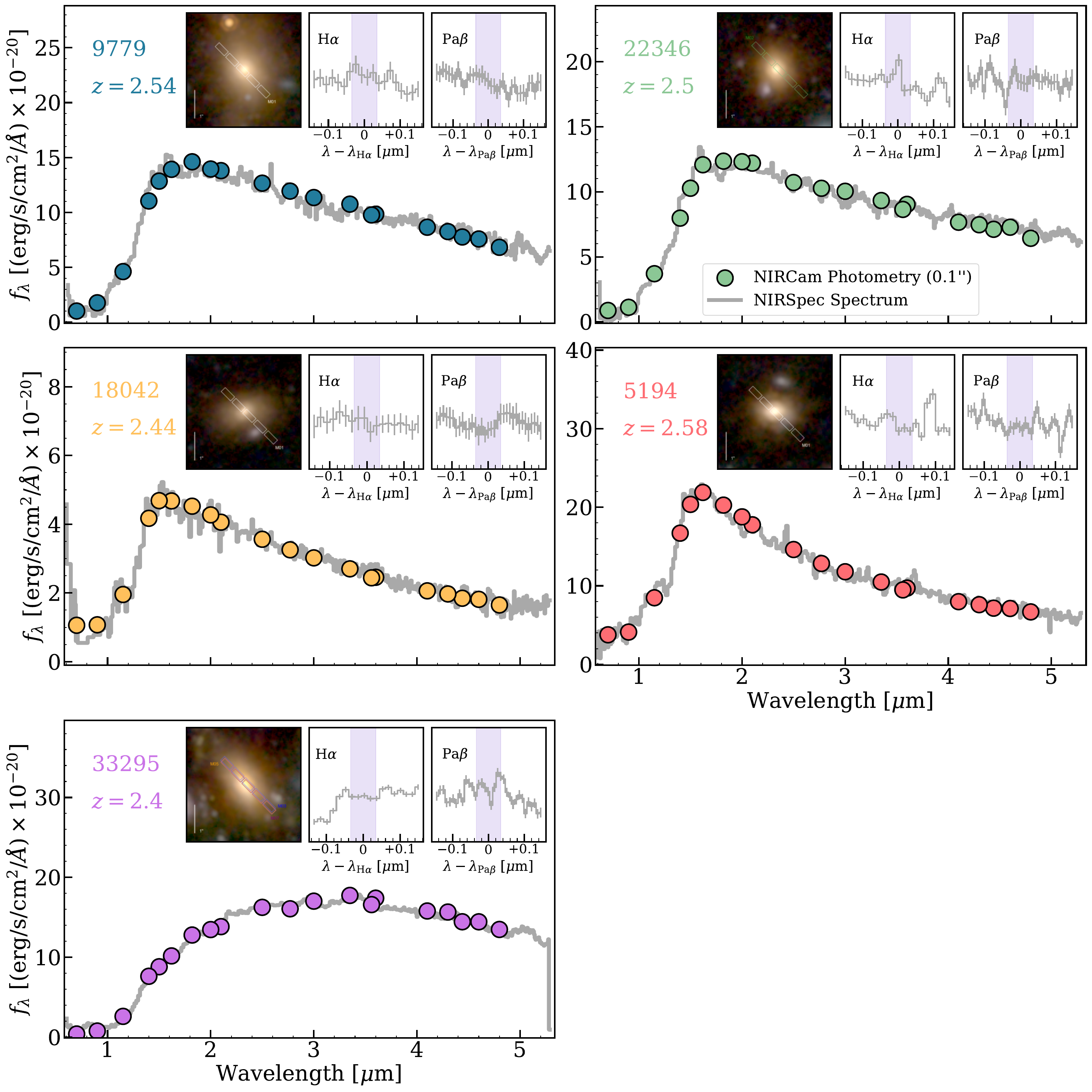}{
\textwidth}{ } }
\caption{
NIRSpec spectra and NIRCam photometry (elliptical annulus of radius $0.11''$) for each target in our sample.
Photometric uncertainties are smaller than the plotted points.
The spectra are flux matched to the photometry.
No target presents strong emission features.
The leftmost inset panel presents RGB images of the galaxies alongside the micro-shutter arrays (MSAs). 
}
\label{fig:collective_spectra}
\end{figure*}

Based on NIRCam imaging, UNCOVER targeted approximately 700 sources for spectroscopic follow up.
The full UNCOVER spectroscopic targeting strategy is discussed in \cite{Price2024}. 
Part of the  targeting strategy was selecting for quiescent galaxies.
Quiescent status was determined by stellar population synthesis fits to F115W$-$F444W NIRCam broadband imaging; F410M medium band and ancillary HST data were considered where available \citep[][Khullar et al. in preparation]{Wang2024}. 
Multiply-imaged galaxies were also prioritized \citep{Furtak2023}.

Of the spectroscopically confirmed galaxies in the UNCOVER DR4 catalog, we select all massive  quiescent  galaxies between $2<z<3$.
Based on the UNCOVER DR3 photometric catalog, we require  $\log_{10}[\frac{M}{M_\odot}]>10$  and  $\log_{10}[\frac{\mathrm{sSFR}}{\mathrm{yr}}]<-10$ (Figure~\ref{fig:collective_selection}).
Five spectroscopically confirmed galaxies meet these criteria. 
In terms of F150W$-$F444W color, the spectroscopically confirmed galaxies span the distribution of the fifteen photometry-only cosmic noon massive quiescent galaxies in the UNCOVER catalog.

The selected galaxies are presented in Table~\ref{tab:light_weighted}.
The redshifts and magnifications are adopted from the UNCOVER DR4 catalog.
All five galaxies reside near $z \sim 2.5$.
The galaxies are well described by 2--dimensional {\sersic} models (see Section~\ref{sec:methods});
the galaxies are largely circular ($b/a\gtrsim0.5$), require {\sersic} indices $\gtrsim2$, and have half-light radii of $\sim1$~kpc.
Figure~\ref{fig:collective_spectra} presents color images of each galaxy.

Our galaxies do not appear to be undergoing interactions or to have any nearby (within $20$~kpc projected distance) bright companions. However, a full investigation of the effect of environment on quenching in the UNCOVER field is underway (Pan et al. in preparation), and we defer work quantifying the large-scale environments of these systems to that work.

All five galaxies are lensed by Abell~2744.
One source---UNCOVER~33295\footnote{Throughout this work, we refer to the galaxies by their internal UNCOVER MSA IDs.}---is triply imaged by the Abell 2744 cluster and was identified as the multiply imaged source number 67 in \cite{Furtak2023}; the images are located at ($3.54429$, $-30.36806$), ($3.54204, -30.37461$), and ($3.54283, -30.37350$).
The first image is minimally distorted ($\mu = 5.9$ with nearly equal shear in the tangential and radial directions).
The other two images are strongly sheared and more dramatically lensed ($\mu = 18.0$ and $10.9$).
NIRSpec observations were obtained for the least distorted image;
unless otherwise stated, we consider this image of UNCOVER~33295.

\newpage
\subsection{Spectra}
\label{sec:obs_spectra}

\begin{deluxetable*}{cccccccc}[t]
\tablecaption{Sample Summary.\label{tab:light_weighted}}
\tablehead{
 \colhead{Target} & \colhead{Ra} & \colhead{Dec} & \colhead{Redshift} & \colhead{Magnification} & \colhead{Half-light radius} & \colhead{Sersic index} & \colhead{Ellipticity} \\
  \colhead{} & \colhead{J2000} &  \colhead{J2000} & \colhead{$z$} & \colhead{$\mu$} & \colhead{[$r_\mathrm{eff}/$kpc]} & \colhead{$n$} & \colhead{$b/a$} 
}
\startdata
9779  & $3.55653$ & $-30.40865$ & $2.54$ & $1.47$ & $2.16_{-0.04}^{+0.04}$ & $4.20_{-0.03}^{+0.03}$ & $0.646_{-0.002}^{+0.002}$\\
22346  & $3.63103$ & $-30.38324$ & $2.50$ & $1.34$ & $1.01_{-0.01}^{+0.01}$ & $2.86_{-0.02}^{+0.02}$ & $0.724_{-0.003}^{+0.003}$\\
18042  & $3.60158$ & $-30.39149$ & $2.44$ & $2.12$ & $1.7_{-0.2}^{+0.2}$ & $6.8_{-0.2}^{+0.2}$ & $0.629_{-0.005}^{+0.004}$\\
5194  & $3.59226$ & $-30.42074$ & $2.58$ & $2.26$ & $0.67_{-0.02}^{+0.02}$ & $2.04_{-0.03}^{+0.03}$ & $0.507_{-0.002}^{+0.002}$\\
33295  & $3.54429$ & $-30.36806$ & $2.40$ & $5.86$ & $0.71_{-0.01}^{+0.01}$ & $2.84_{-0.01}^{+0.01}$ & $0.441_{-0.001}^{+0.001}$\\
\enddata
\tablecomments{The structural parameters are from 2--dimensional {\sersic} fits to each galaxy's F444W image. 
All masses and physical scales are lensing corrected.}
\end{deluxetable*}

\begin{deluxetable*}{cccccccc}[t]
\tablecaption{Inferred Galaxy Properties.\label{tab:results}}
\tablehead{
 \colhead{Target} & \colhead{Stellar Mass} & \colhead{Half-Mass Radius} & \colhead{Surface Density} & \colhead{Age} & \colhead{Quenching Time} & \colhead{Specific SFR}  & \colhead{Dust Mass} \\
  \colhead{} & \colhead{$\log_{10}[ \frac{M}{M_\odot}]$} & \colhead{[$r_\mathrm{eff}/$kpc]} & \colhead{$\log_{10}[ \frac{\Sigma_{<1\mathrm{kpc}}}{M_\odot\mathrm{kpc}^2}]$} & \colhead{$[t_{50}/\mathrm{Gyr}]$} & \colhead{$[t_{90}/\mathrm{Gyr}]$} & \colhead{$\log_{10}[ \frac{\mathrm{sSFR}}{\mathrm{yr}}]$}  & \colhead{$\log_{10}[ \frac{M_\mathrm{dust}}{M_\odot}]$}
}
\startdata
9779 & $11.02_{-0.03}^{+0.03}$ & $1.4_{-0.2}^{+0.3}$ & $10.1_{-0.1}^{+0.1}$ & $1.1_{-0.3}^{+0.2}$ & $0.3_{-0.1}^{+0.1}$ & $-9.9_{-0.2}^{+0.2}$ & $<7.6$\\
22346 & $10.96_{-0.02}^{+0.02}$ & $0.9_{-0.1}^{+0.2}$ & $10.2_{-0.04}^{+0.04}$ & $1.5_{-0.2}^{+0.2}$ & $0.8_{-0.1}^{+0.1}$ & $-10.7_{-0.3}^{+0.3}$ & $<7.6$\\
18042 & $10.23_{-0.02}^{+0.02}$ & $1.1_{-0.2}^{+0.3}$ & $9.41_{-0.05}^{+0.04}$ & $1.3_{-0.2}^{+0.2}$ & $0.5_{-0.1}^{+0.1}$ & $-10.8_{-0.4}^{+0.3}$ & $<7.3$\\
5194 & $10.57_{-0.02}^{+0.02}$ & $0.6_{-0.1}^{+0.1}$ & $9.9_{-0.03}^{+0.03}$ & $1.0_{-0.2}^{+0.3}$ & $0.46_{-0.07}^{+0.04}$ & $-10.4_{-0.3}^{+0.3}$ & $<7.4$\\
33295 & $10.71_{-0.03}^{+0.03}$ & $0.4_{-0.1}^{+0.1}$ & $10.09_{-0.04}^{+0.05}$ & $1.4_{-0.2}^{+0.2}$ & $0.5_{-0.1}^{+0.1}$ & $-10.3_{-0.4}^{+0.4}$ & $<6.5$\\
\enddata
\tablecomments{
The total stellar mass, half-mass radius, and inner $1$~kpc surface density are reported for each galaxy.
The galaxies' mass weighted ages and quenching times (the lookback time by which $50\%$ and $90\%$ of the stellar mass had been formed, respectively) are also presented.
Upper limits ($3\sigma$) on dust mass are inferred from ALMA $1.2$~mm non-detections, see Section~\ref{sec:dust_mass}.
All masses and physical scales are lensing corrected.}
\end{deluxetable*}

\begin{deluxetable}{cccc}[h]
\tablecaption{Constraints on Active Star Formation.\label{tab:sfr}}
\tablehead{
 \colhead{Target} & \colhead{Paschen$-\beta$} & \colhead{FIR Emission} & \colhead{\texttt{Prospector}}\\
  \colhead{} & \colhead{$[M_\odot/\mathrm{yr}]$} & \colhead{$[M_\odot/\mathrm{yr}]$} & \colhead{$[M_\odot/\mathrm{yr}]$}
}
\startdata
9779 & $<14.0$ & $<19.3$ & $13.0_{-4.0}^{+6.0}$\\
22346 & $<4.2$ & $<19.1$ & $2.0_{-1.0}^{+2.0}$\\
18042 & $<4.2$ & $<10.3$ & $0.3_{-0.2}^{+0.3}$\\
5194 & $<0.8$ & $<12.5$ & $2.0_{-1.0}^{+1.0}$\\
33295 & $<6.0$ & $<1.47$ & $3.0_{-2.0}^{+5.0}$\\
\enddata
\tablecomments{
Star formation constraints from multiple tracers. 
The Paschen$-\beta$ and FIR emission upper limits are $3\sigma$.
The \texttt{Prospector} uncertainties are $1\sigma$.
All masses and physical scales are lensing corrected.}
\end{deluxetable}

We only consider spectroscopically confirmed galaxies from the UNCOVER program.
With the galaxies' SEDs well sampled by NIRCam, we solely use the NIRSpec spectra to place constraints on emission line star formation tracers.

The reduction pipeline for the UNCOVER NIRSpec observations is presented in \cite{Price2024}. 
An electrical short occurred during the first UNCOVER NIRSpec pointing, impacting three of our targets: UNCOVER~5194, 9779, and 18042.
As a result, these targets were not processed by \cite{Price2024}.
To ensure uniform data reduction, we perform a set of independent spectroscopic reductions for all our sources \citep[identical to][]{Setton2024}.
First, \texttt{msaexp} reduced stage two data products are retrieved from
MAST \citep[v0.6.10,][]{Brammer2022}; \texttt{msaexp} corrects for known image artifacts and $1/f$ noise.
The observations are next flat fielded and background subtracted;
background spectra are estimated from empty slits.
The spectra are then flux calibrated against the NIRCam photometry.
For each target, we fit a 2nd~order polynomial between synthetic 20--band photometry measured from the spectrum and NIRCam elliptical aperture photometry (radius $0.11''$ with an axis ratio set by a 2--dimensional {\sersic} fit to the galaxy's F444W image, see Section~\ref{sec:methods}).
We use the elliptical aperture photometry, instead of the UNCOVER DR3 catalog photometry, because some of the galaxies present color gradients, and the MSA were approximately aligned with the galaxies' centers.
We rectify the spectra to the photometry by these polynomials; 
the corrections are at the $10\%$ level.

All five spectra lack strong emission features.
Figure~\ref{fig:collective_spectra} presents the flux calibrated spectra alongside the aperture photometry.
Assuming the UNCOVER reported spectroscopic redshifts (Table~\ref{tab:light_weighted}), none of the galaxies present detectable emission features ($3\sigma$), although UNCOVER~22346 and 5194 show hints of line emission in the vicinity of H$\alpha$.
To constrain the galaxies' ongoing star formation rates, we place upper bounds on each galaxies' H$\alpha$ and Paschen--$\beta$ flux in Section~\ref{sec:sfh}.

\section{Spatially Resolved Characterization}
\label{sec:methods}

\begin{figure*}[t]
\gridline{\fig{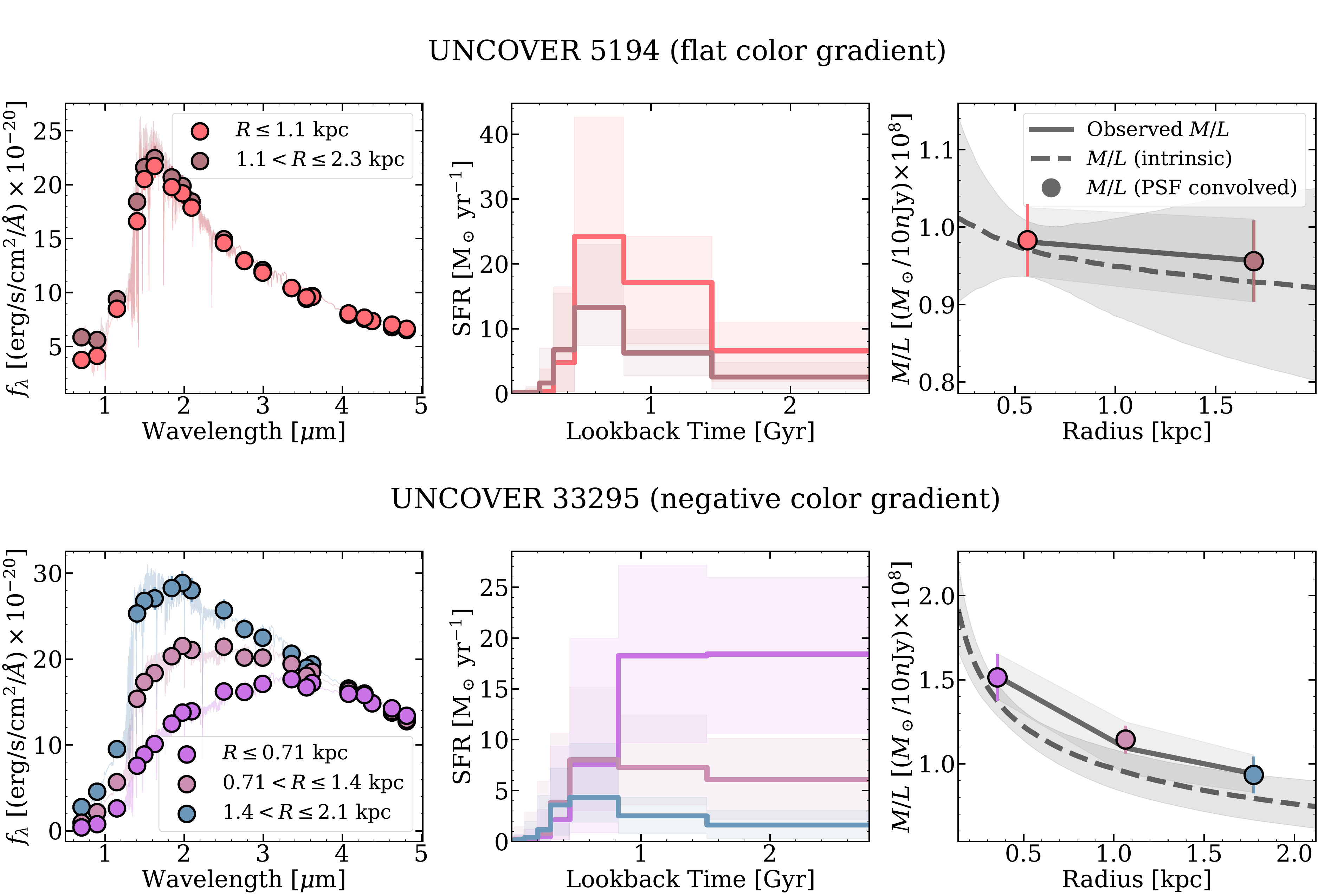}{
\textwidth}{ } }
\caption{
For two representative galaxies---UNCOVER~5194 (flat color gradient) and 33295 (negative color gradient)---we diagram the annular SED fitting process. 
The three other galaxies in our sample are presented in Appendix~\ref{sec:spatially_resolved_appendix}.
Left: the annular SEDs (matched to the central annuli's F444W flux); 
the \texttt{Prospector} fits are shown alongside the annular SEDs.
Center: the inferred star formation histories for each annulus.
Right: the observed mass-to-light $M/L$ ratio for each annulus (solid line); 
the intrinsic $M/L$ profiles  (inferred from forward modeling) are shown as dashed lines, with the corresponding PSF convolved $M/L$ ratios presented as circles.
The shaded regions and errorbars demarcate $1\sigma$ uncertainties.
}
\label{fig:methods}
\end{figure*}
Each target was observed with all NIRCam medium- and wide-band filters (20 bands total, see Figure~\ref{fig:collective_spectra}),
corresponding to $0.7-4.8~${\microns} in observed-frame and $0.2-1.4~${\microns} in rest-frame ($z\sim 2.5$).
To fully leverage this dense wavelength coverage and JWST's high angular resolution, we perform spatially resolved stellar population synthesis modeling.
Our approach closely follows \cite{Suess2019a} and \cite{Setton2024}, which we describe briefly below.

\subsection{Annular Photometry}
\label{sec:annular}

To characterize the structure and stellar populations of the sample, we investigate how the galaxies' SEDs vary with radius.
Throughout this work, we adopt the empirical PSFs of \cite{Weaver2024} and resolution match all images to the F444W PSF.

First, we model each target's F444W image as a single component 2--dimensional {\sersic} profile with \texttt{pysersic} \citep{Pasha2023}.
The best-fit {\sersic} parameters are presented in Table~\ref{tab:light_weighted}.
We next extract spatially resolved SEDs from mosaic images that have been PSF-matched to the lowest resolution images (F444W) using elliptical annuli.
The annuli's axis ratios are adopted from the 2--dimensional {\sersic} fits; to account for the PSF, the intrinsic axis ratio from the {\sersic} fit ($q_\mathrm{intrinsic}$) is related to the PSF convolved axis ratio ($q_\mathrm{PSF~convolved}$) following \cite{Suess2019a}:
\begin{equation}
    q_\mathrm{PSF~convolved} = \sqrt{\frac{(q_\mathrm{intrinsic} r_\mathrm{eff})^2 + r_\mathrm{PSF~FWHM}^2 }{r_\mathrm{eff}^2 + r_\mathrm{PSF~FWHM}^2}},
\end{equation}
where $r_\mathrm{eff}$ is the F444W semi-major axis half-light radius and $r_\mathrm{PSF~FWHM}$ is the F444W PSF FWHM.
For the $n$th annulus (where $n=1,2,\dots$), the inner radius is fixed to $(n-1) \times r_\mathrm{PSF~FWHM}$ with an outer radius of $n \times r_\mathrm{PSF~FWHM}$; 
$r_\mathrm{PSF~FWHM}$ is set as the radial step-size to ensure the annuli are effectively independent of one another.
Residual sky flux is treated as the median of all pixels not in the segmentation map \citep{Weaver2024}, within a $5.6'' \times 5.6''$ box around the target (i.e., $140 \times 140$ pixels).
To avoid low signal-to-noise ratios, annuli are limited to radii less than three half-light radii $r_\mathrm{eff}$.

\subsection{Stellar Population Synthesis}
\label{sec:pop_synth}

\begin{figure*}[t]
\gridline{\fig{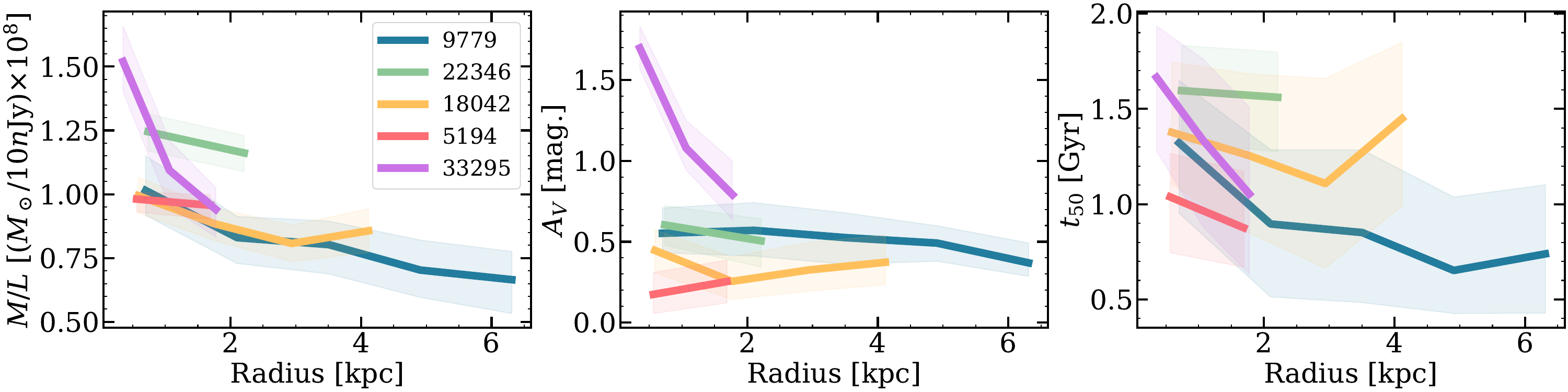}{
\textwidth}{ } }
\caption{ Inferred radial gradients from \texttt{Prospector} fits to annular photometry.
The galaxies predominantly display negative color gradients (redder core) with non-zero dust attenuation in both the core and outskirts.
From left to right: mass-to-light ratio $M/L$, reddening $A_V$, and mass-weighted age $t_{50}$.
The median and $1\sigma$ confidence interval for a given parameter are shown as a solid line and shaded region, respectively. 
}
\label{fig:collective_gradients_truncated}
\end{figure*}

To map the SEDs to stellar populations, we employ \texttt{Prospector}, a toolkit for Bayesian  population synthesis \citep{Johnson&Leja2017, Leja2017, Johnson2021}.
As inputs, we adopt the Flexible Stellar Population Synthesis (FSPS) models \citep{Conroy2009,Conroy2010}, the MILES spectral library \citep{Sanchez-Blazquez2006}, MESA Isochrones and Stellar Tracks \citep[MIST,][]{Choi2016,Dotter2016}, and the \cite{Chabrier2003} initial mass function.
We model star formation histories (SFHs) non-parametrically with eight bins. 
The first five SFH bins have widths of $20$, $100$, $200$, $300$, and $450$~Myr; the remaining lookback time is filled by three logarithmically distributed bins.
Following \cite{Leja2019}, continuity in the star formation histories is encouraged via a Student's~T ($\nu=2$) prior of width $0.3$ on the logarithmic difference in star formation rate  between adjacent bins (i.e., $\log_{10} \mathrm{SFR}_{i+1} - \log_{10} \mathrm{SFR}_i$).
Dust attenuation follows \cite{Kriek2013} with amplitude $A_V\in (0,2.5)$ and power-law index $\delta \in (-1,0.4)$ that acts multiplicatively on a \cite{Calzetti2000} dust law;
the attenuation of young stars ($<10$~Myr) is fixed to twice the attenuation of the older stars \citep{Wild2020}.
We treat metallicity as a free-parameter, with a uniform prior of $Z/Z_\odot\in(0.01,1.5)$.

Each annular SED is modeled independently.
Parameter inference is conducted via \texttt{dynesty} nested sampling \citep{Speagle2020,Koposov2022}.
The signal-to-noise ratio of each band is capped at $20$ to reflect additional sources of uncertainty \citep[e.g., flux calibration and model systematics, see][]{Wang2024}.
Our procedure for spatially resolved stellar characterization is diagrammed in Figure~\ref{fig:methods};
for UNCOVER~5194 (no color gradient) and 33295 (strongly negative color gradient), we present the SED of each annuli, alongside the best fitting \texttt{Prospector} model and the inferred star formation history (the three other galaxies in our sample are presented in Appendix~\ref{sec:spatially_resolved_appendix}).

\subsection{Mass-weighted structure}
\label{sec:structure}

Unless the mass-to-light ratio of a galaxy is spatially uniform, its light-weighted properties (e.g., half-light radius and {\sersic} index) offer a biased view of the underlying stellar mass distribution (particularly when considering bluer bands).
We leverage our spatially resolved population synthesis results to reconstruct each galaxy's stellar mass distribution.
Our approach follows \cite{Suess2019a}, which we briefly outline below.

As described above, each galaxy is divided into elliptical annuli and independently modeled with \texttt{Prospector}; 
this yields a mass-to-light ($M/L$) radial profile for each target. 
However, these $M/L$ gradients are PSF-convolved.
To infer a galaxy's intrinsic $M/L$ gradient, we adopt a forward modeling approach.
A given $M/L$ gradient---assumed to take the form $\log_{10}(M/L)=\alpha \times \log_{10}(R) + \beta$ \citep{Chan2016,Suess2019a}---is convolved with the F444W PSF and then compared with the $M/L$ ratios inferred by \texttt{Prospector}.
By repeating this procedure with MCMC sampling \citep{ForemanMackey2013}, we infer the intrinsic $M/L$ profile for each galaxy.
Multiplying these $M/L$ profiles with each galaxy's 2--dimensional F444W {\sersic} fit yields the underlying mass distributions.
This process is diagrammed in Figure~\ref{fig:methods}.

\section{Results}
\label{sec:results}

Using all 20 medium- and wide-band NIRCam filters, we perform spatially resolved stellar population synthesis on a sample of five quiescent UNCOVER galaxies ($z \sim 2.5$).

The galaxies predominantly display negative color gradients (redder core and bluer outskirts).
One galaxy (UNCOVER~33295) has a strongly negative $M/L$ gradient ($3\sigma$), while most of the sample are weakly negative ($1\sigma$); only UNCOVER~5194 is consistent with flat.
For the majority of the sample (4/5), the galaxies' half-mass radii are therefore smaller than their F444W half-light radii.
These color gradients are typical of massive quiescent galaxies at cosmic noon \citep{Miller2022,Suess2022}. 

The global properties of the galaxies are presented in Table~\ref{tab:results}, including their mass-weighted structures, characteristic ages, and active star formation rates.
The inferred structures are typical of $z\sim2.5$ massive quiescent galaxies: stellar masses between $10^{10}$ and $10^{11}~M_\odot$ with half-mass radii between $0.5$ and $2.0$~kpc \citep[see][]{Suess2019b}.

\begin{figure*}[t]
\gridline{\fig{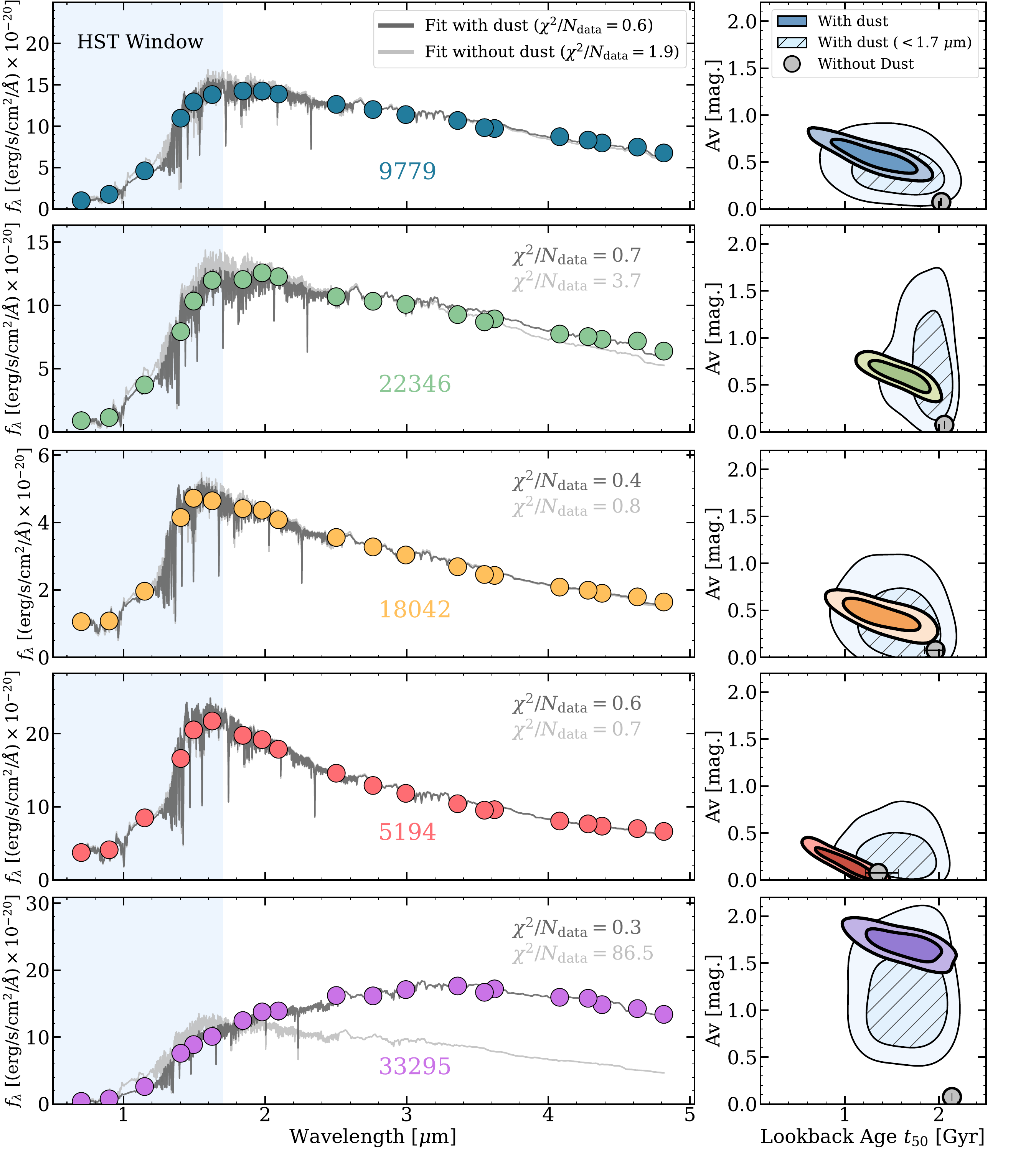}{
\textwidth}{ } }
\caption{
The majority of our galaxies are inconsistent with dust-free stellar populations.
Left: each galaxy's observed SED (central annuli) compared to the best fit \texttt{Prospector} model, either with (dark-grey) or without (light-grey) dust attenuation as a free-parameter;
goodness of fit (in terms of $\chi^2$ divided by the number of data points) is reported for both models.
Photometric uncertainties are smaller than the plotted points.
The approximate wavelength range accessible to HST is shaded light-blue.
Right: the posteriors on dust attenuation $A_V$ versus age $t_{50}$ for \texttt{Prospector} models with and without dust attenuation. 
To highlight the power of NIRCam imaging, posteriors are also presented for a dusty \texttt{Prospector} model fit only to the $\lambda<1.7$~{\microns} photometry.
The shaded contours represent the $50$ and $84$th percentiles.
The posteriors for the dust-free models are shown as grey circles near $A_V\sim0$, with errorbars representing the $1\sigma$ confidence intervals on $t_{50}$. 
} 
\label{fig:dust_or_no_dust}
\end{figure*}

\subsection{Tracers of Quiescent Status}
\label{sec:sfh}

Multiple tracers confirm the galaxies' quiescent status.
The NIRSpec spectra present minimal line emission; Paschen--$\beta$ and H$\alpha$ are undetected ($3\sigma$) for all five galaxies (see Figure~\ref{fig:collective_spectra}).
We set an upper limit on each galaxy's Paschen--$\beta$ flux by modeling the observed spectrum as a Gaussian, with the continuum treated as a linear function of wavelength (optimized via MCMC sampling).
The strength of dust attenuation is adopted from each galaxy's central annuli \texttt{Prospector} fit (since the MSA are approximately aligned with the galaxies' centers).
Upper bounds ($3\sigma$) on Paschen--$\beta$ flux are converted to SFR via Eqn.~2 of \cite{Reddy2023};
presented in Table~\ref{tab:sfr}.

For H$\alpha$, this process is complicated by two factors: (i) at NIRSpec's resolution H$\alpha$ and the [NII] doublet are blended (and lie in close proximity to the [SII] doublet) and (ii) H$\alpha$ absorption is non-negligible. 
Assuming H$\alpha$ dominates the line emission---and correcting for both stellar absorption and dust attenuation---the H$\alpha$ SFR limits are comparable to the Paschen--$\beta$ constraints.
Given the greater systematic uncertainties for H$\alpha$, we opt to only report the Paschen--$\beta$ limits.

All targets are also undetected in the $1.2$~mm data from the DUALZ ALMA survey \citep{Fujimoto2023}. 
Assuming FIR dust emission ($T=35$~K with a spectral index of $1.8$), each target's $3\sigma$ upper limit on $1.2$~mm emission is mapped to an ongoing star formation rate (Table~\ref{tab:sfr}).
The FIR SFR limits are considerably stronger for UNCOVER~33295 (a triply imaged source).
The most strongly lensed image ($\mu \sim 18$) sets a FIR SFR limit of $\lesssim 1.5$~{\MsolPerYr}.

Based on SED shape, the \texttt{Prospector} models also favor quiescent solutions, consistent with the Paschen--$\beta$ and FIR bounds.
Across all star formation tracers, each galaxy requires a specific star formation rate of $\log_{10}[ \frac{\mathrm{sSFR}}{\mathrm{yr}}] \lesssim -10$, i.e., $1.5$~dex below the star forming main sequence at $z\sim2.5$ \citep{Leja2022}.

\subsection{Dusty through and through}
\label{sec:dusty_results}

For each galaxy, we report mass-to-light ratio ($M/L$), dust attenuation ($A_V$), and age ($t_{50}$,  the lookback time by which half the stellar mass had been formed) as a function of radius; see Figure~\ref{fig:collective_gradients_truncated}.
Metallicity is poorly constrained for all five galaxies, but is marginalized over in order to account for its covariances with age and dust. 
These profiles are PSF convolved, unlike the intrinsic $M/L$ profiles inferred by forward modeling in Section~\ref{sec:structure}.
The galaxies typically present negative color gradients and favor dusty solutions;
UNCOVER 33295 has particularly strong radial gradients (see Section~\ref{sec:33295}).
All five galaxies have central dust attenuation levels of $A_V \gtrsim 0.2$, with most favoring $A_V \gtrsim 0.4$.
Dusty solutions are also favored in the outskirts of the galaxies.
The inferred stellar populations formed $\sim1$~Gyr ago in lookback time (corresponding to a formation redshift of $z\sim4$), with quenching times of $t_{90} \sim 0.5$~Gyr. 
Taken together, these results reveal dusty quiescent galaxies with negative color gradients.

All five galaxies favor dusty stellar populations. 
However, interpreting these solutions is complicated by the dust--age degeneracy in stellar population synthesis.
Fortunately, the broad wavelength coverage of JWST NIRCam  significantly suppresses this degeneracy.
To quantify this improvement, we again fit each galaxy's annular SEDs with \texttt{Prospector} but only for bands blueward of $1.7$~{\microns} (the approximate wavelength range of HST).  
Figure~\ref{fig:dust_or_no_dust} presents the posteriors on reddening and stellar age for our fiducial fits (all 20 bands) and the blue--only fits ($\lambda < 1.7$~{\microns}).
The constraints on dust attenuation and mass-weighted age are consistent at 1$\sigma$ between the two fits---i.e., the exclusion of the reddest bands does not systematically skew the posteriors.
For all five galaxies, the red NIRCam bands dramatically improve our constraints on reddening and stellar age.
The improvement is particularly pronounced for the reddening measurements;
without the reddest filters, only one galaxy (UNCOVER~33295) favors non-zero reddening, while at least three galaxies favor significant reddening when all 20 bands are used.

To explore the significance of the dusty solutions, we next consider whether the galaxies are consistent with dust-free stellar populations.
We remodel the annular SEDs with reddening fixed to $A_V=0$.
Figure~\ref{fig:dust_or_no_dust} presents the observed SEDs for the galaxies' central annuli alongside the best fitting dusty and dust-free solutions.
The dust-free models are naturally less flexible and thus fit the data only as well or worse than the models with dust attenuation.
For UNCOVER 22346 and 33295, dust-free stellar populations clearly fail to reproduce the observed SED. Figure~\ref{fig:dust_or_no_dust} presents each model's posterior on dust attenuation $A_V$ versus age $t_{50}$, as well as a goodness-of-fit metric ($\chi^2$ divided by the number of data points).
Only for UNCOVER~5194 (the target with the lowest central dust attenuation $A_V\sim0.2$) do the posteriors of the dusty and dust-free models overlap (at the $1\sigma$ level); the dusty and dust-free posteriors are also similar for UNCOVER~18402.
The dust-free models also require extreme star formation histories.
Without dust, most of our sample (3/5) would require maximally old stellar populations.
We therefore favor attenuated stellar populations over maximally old dust-free solutions for the majority of our sample.

\section{Discussion}
\label{sec:discussion}

We report a sample of five massive ($\log_{10}[M / M_\odot] \sim 10-11$) quiescent galaxies at $z \sim 2.5$.
We use all 20 medium- and wide-band NIRCam filters, as well as NIRSpec and ALMA coverage. 
Our investigation reveals four key results:
(i) the galaxies are indeed quiescent, with all five satisfying $\log_{10}[ \frac{\mathrm{sSFR}}{\mathrm{yr}}] \lesssim -10$ via numerous tracers of the star formation rate; (ii) the galaxies' mass-weighted structures (e.g., total stellar mass, half-mass radius, and central surface density) are consistent with the larger population of $z \sim 2.5$ quiescent galaxies; (iii) the galaxies are notably reddened: all five favor dust attenuation levels of $A_V\gtrsim0.2$, and all but one favor $A_V\gtrsim0.4$; and (iv) most of the galaxies (4/5) display weak negative color gradients 
(the exception is UNCOVER~33295, which has strong $M/L$ and $A_V$ gradients).

In the local Universe, quiescent galaxies are typically dust poor.
Our sample of dusty quiescent galaxies at $z \sim 2.5$ implies a decline in dust content from cosmic noon to the present. 
To contextualize our results, we build a timeline of quiescent galaxy reddening from $z \sim 0$ to $2.5$, in Section~\ref{sec:dusty_disc}.
We consider the galaxies' dust masses (from FIR emission) in Section~\ref{sec:dust_mass}.
In Section~\ref{sec:33295}, we discuss the implications of UNCOVER~33295's radial $A_V$ gradient, as well the color gradients of similar quiescent galaxies.

\subsection{Dusty quiescent galaxies}
\label{sec:dusty_disc}

\begin{figure}[t]
\gridline{\fig{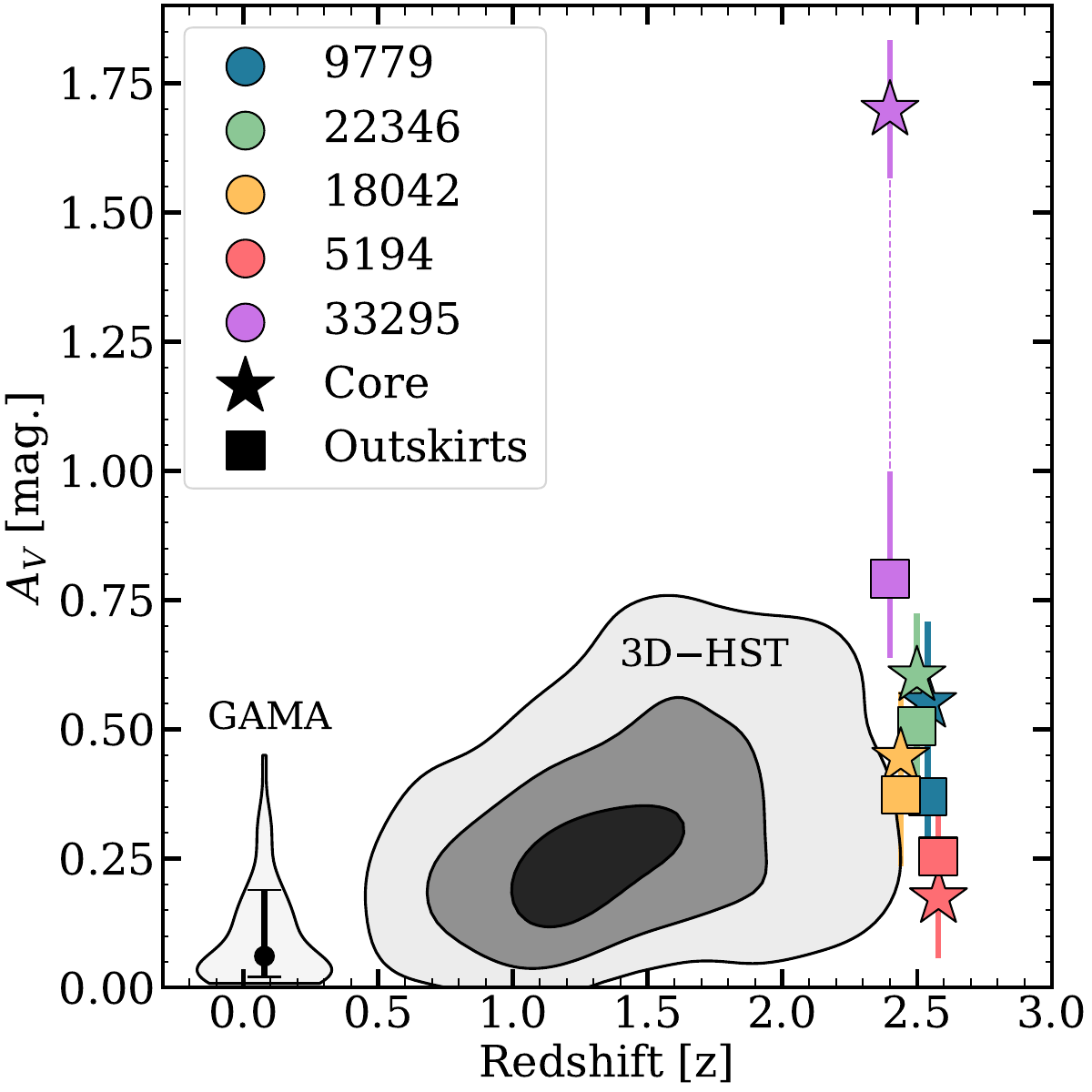}{\columnwidth}{ } }
\caption{
The typical reddening of massive ($\log_{10}[ \frac{M}{M_\odot}]\gtrsim10$) quiescent galaxies declines between $z\gtrsim2$ and the present.
For each galaxy in our $z\sim 2.5$ sample, we present the inferred dust attenuation $A_V$ for the inner- and outermost annuli.
The evolution of $A_V$ with redshift for massive quiescents in the 3D--HST sample is presented in grey \citep{Suess2019b};
the contours correspond to the $16, 50$, and $84$th percentiles.
The distribution of $A_V$ for $z<0.1$ massive quiescent galaxies in the GAMA survey is shown in grey; the central point and errorbars correspond to the median and $1\sigma$ intervals.
}
\label{fig:collective_Av_vs_z}
\end{figure}

\begin{figure*}[t]
\gridline{\fig{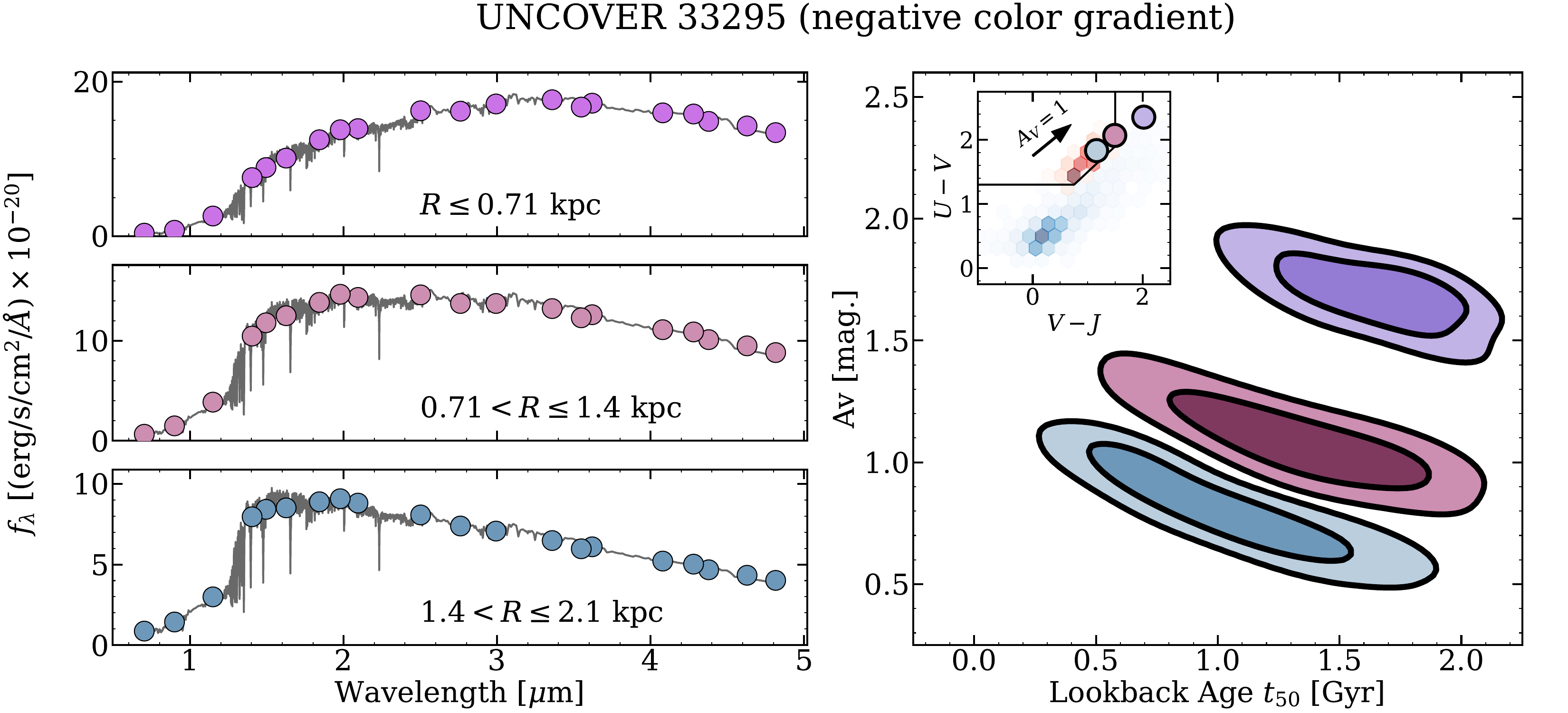}{
\textwidth}{ } }
\caption{
Demonstration of the negative color gradient of UNCOVER~33295. 
Left: annular SEDs alongside their corresponding \texttt{Prospector} models (grey lines).
Photometric uncertainties are smaller than the plotted points.
Right: posteriors on dust attenuation $A_V$ and age $t_{50}$ for the \texttt{Prospector} models; 
the contours correspond to the $50$ and $84$th percentiles.
The inset panel presents the rest frame $V-J$ and $U-V$ colors for the three annuli, alongside the $2<z<2.5$  star forming (blue shaded) and quiescent (red shaded) galaxies in the 3D--HST field \citep{Suess2019b}.
} 
\label{fig:33295_colors}
\end{figure*}
Dust attenuation and stellar age are notoriously degenerate, even more so without wide wavelength coverage.
Previous studies of $z \gtrsim 1$ quiescent galaxies were often unable to confidently distinguish between dust reddening and stellar age.
In this paper, we lift the dust--age degeneracy by leveraging all 20 medium- and wide-band NIRCam filters (rest-frame $0.2-1.4~${\microns}).
We conclude that our sample of five $z \sim 2.5$ galaxies are unambiguously quiescent, with the majority inconsistent with dust-free stellar populations.

Prior studies of $z \gtrsim 1$ quiescent galaxies have hinted at significant dust reddening.
With NIR photometric coverage from UltraVISTA and ZFOURGE photometry, \cite{Marchesini2014} and \cite{Straatman2016} both found a gradual decline in the typical $A_V$ of quiescent galaxies with cosmic time (from $0.5$ at $z \sim 2$ to near zero in the local Universe).
\cite{Martis2019} found that a substantial fraction of reddened galaxies ($A_V\gtrsim1$) between $1 \lesssim z \lesssim 4$ are indeed quiescent, using UltraVISTA and Herschel photometry. 
More recently, JWST has detected several reddened quiescent galaxies at $z \gtrsim 3$ \citep{Rodighiero2023,Alberts2023,Nanayakkara2024,Setton2024}. 

Mounting evidence for the existence of reddened quiescent galaxies at $z \gtrsim 1$ implies a decline in the typical $A_V$---and therefore, the ISM content---of quiescent galaxies with cosmic time.
For context, we present two comparison samples: (i) the \cite{Suess2019b} sample of massive quiescent galaxies in 3D--HST ($0.5 \lesssim z \lesssim 2.5$) and (ii) low redshift ($z < 0.1$) massive quiescent galaxies from the GAMA survey.
The selected galaxies are massive ($\log_{10}[ \frac{M}{M_\odot}] \gtrsim 10)$ and satisfy the rest-frame $UVJ$ quiescence criteria  \citep{Whitaker2011}.
Stellar population fits are publicly available for both samples \citep[e.g.,][]{Taylor2011,Straatman2016}.
For internal consistency, we perform our own \texttt{Prospector} modeling of the galaxies' SEDs; see Appendices~\ref{sec:appendix_3dhst} and \ref{sec:appendix_gama} for descriptions.

The evolution of $A_V$ with redshift for massive quiescent galaxies is presented in Figure~\ref{fig:collective_Av_vs_z}.
The inferred $A_V$ for our galaxies' inner- and outermost annuli are shown, alongside the two comparison samples: 3D--HST ($0.5 \lesssim z \lesssim 2.5$) and GAMA ($z<0.1$).
The $A_V$ distribution at a given redshift naturally marginalizes over galaxies' many geometric orientations and formation pathways. 
Even at low redshift---where quiescent galaxies are predominantly unattenuated---some quiescent galaxies host prominent dust lanes and/or centrally concentrated dust, likely resulting from recent mergers \citep[e.g.,][]{Ebneter1985, Ebneter1988, vanDokkum1995, Tomita2000, Tran2001, Lauer2005,Whitaker2008, Davis2013, Boizelle2017}.
Our focus is the typical $A_V$ at a given redshift---i.e., how uncommon are quiescent galaxies that are as dusty as our $z\sim2.5$ sample at different cosmic times?
Between cosmic noon and the present, the typical reddening falls from $A_V \sim 0.5$ to near 0. 
The low redshift GAMA galaxies have a median reddening of $A_V \sim 0.06$, with a $95$th percentile of $A_V \sim 0.3$; in contrast, all but one of our UNCOVER galaxies requires  $A_V \gtrsim 0.4$.
Our five  $z\sim 2.5$  galaxies would be among the most reddened quiescent galaxies in the local Universe. 

Is our sample of quiescent galaxies biased to the reddest and thus dustiest sources?
Our sample of five spectroscopically confirmed galaxies are not systematically redder than the ``photometry only" UNCOVER sample of cosmic noon massive quiescent galaxies.
In terms of F150W$-$F444W and rest-frame $V-J$ versus $U-V$, the spectroscopically confirmed galaxies span the color distribution of the fifteen photometry only galaxies.
We therefore conclude our sample is representative of massive quiescent galaxies in UNCOVER and not strongly biased toward dusty galaxies.

If we imagine galaxies maintain an equilibrium between star formation, gas mass, and mass loss, quiescent galaxies are expected to be gas depleted \citep{Dave2012,Lilly2013}.
While quiescent galaxies are indeed dust-poor in the local Universe, our picture of galaxy evolution must now include the decline in galaxy reddening with cosmic time, even in systems no longer actively forming stars.
Recently, several studies have investigated the evolution of quiescent galaxies' gas reservoirs; 
these works were largely motivated by new constraints on the dust content of quiescent galaxies---and hints of the typical dust fraction falling with cosmic time \citep{Gobat2018,Martis2019,Whitaker2021}---from observations in the far-IR to radio.
Assuming the rate of ISM depletion is proportional to the ISM mass, \cite{Gobat2020} interpret the decline in ISM content of quiescent galaxies with cosmic time as a progenitor effect: at higher redshifts, the typical gas and dust fraction of quenched galaxies is propped up due to the relatively recent quenching events.

Hydrodynamical simulations tell a more complex story. 
\texttt{SIMBA} simulations predict considerable diversity in the dust content of quenched galaxies, including wide ranging gas to dust ratios \citep{Whitaker2021simba};
\cite{Lorenzon2024} attributes the varied efficiency of dust depletion in \texttt{SIMBA} quiescent galaxies to the relative timescales of AGN feedback and dust growth. 
With interesting implications for our results, simulations also suggest the depletion timescales for dust are longer than for the cold gas \citep{Lorenzon2024}.
Testing these latest galaxy evolution models will require a large sample of quiescent galaxies at cosmic noon---e.g., how does $A_V$ depend on stellar age, environment, and metallicity?
With rest-frame coverage of $0.2-1.4~${\microns} at $z \sim 2.5$, NIRCam can minimize the dust--age degeneracy and is well poised to construct the large sample of quiescent galaxies at $z \gtrsim 1$ required to address this multi-dimensional question.

\subsection{Dust Masses}
\label{sec:dust_mass}

Our sample of dusty quiescent galaxies at cosmic noon implies a decline in the typical dust content of passive galaxies with cosmic time.
Motivated by this result, we consider an additional probe of our galaxies' dust content: FIR emission.
All five galaxies are undetected in the ALMA $1.2$~mm  continuum from the DUALZ survey \citep{Fujimoto2023}. 
We are therefore limited to placing upper bounds on the galaxies' dust masses.
These nondetections are not in tension with our reddening measurements; 
dust masses consistent with our nondetections have been measured for galaxies with similar levels of reddening  \citep[e.g., Figure 9 of][]{Hodge2024}.
Mapping between $A_V$ and dust mass is non-trivial and highly variable; different star to dust geometries and projection angles can induce significant scatter in $A_V$ for a given dust mass.

We map the $3\sigma$ limits on $1.2$~mm emission to dust masses following Eqn.~2 of \cite{Greve2012}, assuming cold dust ($T=35$~K).
The upper limits on dust mass are reported in Table~\ref{tab:results}.
All five galaxies are constrained to $\log_{10} (M_\mathrm{dust} / M_\odot) \lesssim 7.5$, corresponding to dust fractions of $M_\mathrm{dust} / M_\star \lesssim 10^{-3}$.
Analogous to the FIR SFR limits, the dust mass constraint is strongest for UNCOVER~33295 (a triply imaged source); 
the most strongly lensed image ($\mu \sim 18$) sets a FIR dust mass limit of $\log_{10} (M_\mathrm{dust} / M_\odot) \lesssim 6.5$ and $M_\mathrm{dust} / M_\star \lesssim 10^{-4}$.

Previous FIR observations demonstrate that quiescent galaxies are indeed less dust enriched than star forming galaxies \citep[by at least two dex in the local Universe and at cosmic noon,][]{Magdis2021,Whitaker2021}.
However, inferring the dust content of quiescent galaxies is challenging at $z\gtrsim1$; quiescent galaxies’ lower dust fractions correspond to weaker FIR emission, and the current measurements show considerable scatter.
From co-adding non-detections, \cite{Gobat2018} and \cite{Magdis2021} found the dust fraction of quiescent galaxies declines from $10^{-3}$ at $z\sim2$ to $10^{-5}$ in the local Universe.
\cite{Whitaker2021} constrained the dust masses of six strongly lensed quiescent galaxies ($z\sim 1.5$ to $3$) with ALMA $1.2$~mm observations;
all six galaxies require  $M_\mathrm{dust} / M_\star \lesssim 10^{-3}$, with $4/6$ satisfying  $M_\mathrm{dust} / M_\star \lesssim 10^{-4}$.

Our dust attenuation measurements motivate future observations of the dust continuum in reddened quiescent galaxies to constrain the masses and temperatures of their dust reservoirs and study what drives dust destruction and/or removal.

\subsection{Color gradients: UNCOVER 33295}
\label{sec:33295}

The structures of quiescent galaxies offer valuable insight into their quenching channels, accretion histories, and gas reservoirs.
For the majority of our sample ($4/5$), the galaxy's inferred $M/L$ profile is weakly negative (redder core) or flat (Figure~\ref{fig:collective_gradients_truncated}).
Only UNCOVER 33295 presents strong radial gradients; its mass-to-light ratio declines by nearly $50\%$ from the core to the outskirts, and the radial gradient $\frac{d}{dR} (M/L) = -0.4_{-0.1}^{+0.1}\times10^8~[(M_\odot / 10n\mathrm{Jy} / \mathrm{kpc})]$ is non-zero at the $3\sigma$~level.
The $M/L$ gradient results from high central dust attenuation ($A_V\sim1.5$ in the core and $\sim1$ in the outskirts);
the dust attenuation gradient is non-zero at the $3\sigma$~level $\frac{d}{dR} (A_V)= -0.7_{-0.2}^{+0.2}~[  \mathrm{mag}. / \mathrm{kpc}]$, while the age gradient is consistent with zero $\frac{d}{dR} (t_{50}) = -0.4_{-0.4}^{+0.4}~[\mathrm{Gyr}/\mathrm{kpc}]$.

UNCOVER~33295 is triply-imaged, with $\mu\gtrsim6$ for all three images.
For our photometric and spectroscopic analysis, we consider the least distorted image---i.e., near equal radial and tangential shear ($\mu = 5.9$).
UNCOVER~33295 is the most strongly lensed member of our sample, which raises the question: is UNCOVER~33295 physically unique among our sample or is this galaxy simply better resolved spatially?
We repeat the annular photometry extraction and modeling procedure (Section~\ref{sec:methods}) with the annuli sizes scaled to mimic an intrinsic magnification of $\mu=1.5$, rather than the observed value of $\mu=5.9$.
With the ``less magnified'' annuli, \texttt{Prospector} still favors $A_V\gtrsim1$ in the core. 
The gradients in mass-to-light ratio $\frac{d}{dR} (M/L) = -0.4_{-0.2}^{+0.2} \times10^8~[(M_\odot / 10n\mathrm{Jy} / \mathrm{kpc})]$ and dust attenuation $\frac{d}{dR} (A_V)= -0.4_{-0.1}^{+0.2}~[  \mathrm{mag}. / \mathrm{kpc}]$ are still stronger than any of the other galaxies' radial gradients.
Naturally, the larger annuli decrease the gradients' strengths, however, the ``demagnified'' UNCOVER~33295 still presents detectable radial dependence.
The radial structure of UNCOVER~33295 could therefore have been detected for the other members of our sample.

At $z \gtrsim 1$, a growing number of quiescent galaxies display dust induced negative color gradients (i.e., red interior). 
\cite{Setton2024} recently reported a $z=3.97$ massive quiescent galaxy with a strong $M/L$ gradient (UNCOVER~18407).
Similar to our target (UNCOVER~33295),
the $M/L$ gradient results from high central dust attenuation; with \texttt{Prospector}, \cite{Setton2024} inferred reddening levels of $A_V\sim1.5$  and $0.2$ in the core and  outskirts, respectively.
The mass-weighted structures of UNCOVER~18407 and 33295 are qualitatively similar.
Both quiescent galaxies are massive ($\log_{10}[M/M_\odot]=10.34$~and~$10.71$, respectively), compact ($\log_{10}[\Sigma_{<1\mathrm{kpc}}/M_\odot\mathrm{kpc}^{-2}]=9.61$~and~$10.09$, respectively), and consistent with the distribution of $z\sim2.5$ quiescent galaxies in 3D--HST \citep{Suess2019b}.
Several other cosmic noon galaxies display similar color gradients but lack conclusive population synthesis.
From CEERS NIRCam photometry, \cite{Miller2022} investigated the rest-frame $UVJ$ colors gradients of $>10$~quiescent galaxies between $1.5 < z < 2.5$. 
Over half of the sample displays color gradients, spanning a wide range of magnitudes and directions.
The physical source of these gradients is unclear; 
for quiescent galaxies, dust and age are highly degenerate in $UVJ$ space.
While the occurrence rate of galaxies similar to UNCOVER~33295 remains uncertain, these galaxies have proven relatively common at $z \gtrsim 1$.

\section{Conclusions}
\label{sec:future}
We studied five massive quiescent galaxies at $z \sim 2.5$.
Multiple tracers confirm the galaxies are indeed quiescent, including a lack of $H \alpha$, Paschen--$\beta$, and FIR emission. 
With rest-frame photometric coverage of $0.2-1.4~${\microns}, we suppressed the dust--age degeneracy and concluded the galaxies are predominately dusty;
the majority favor $A_V\gtrsim0.4$ and are incompatible with dust-free stellar populations.

Our sample implies an evolution in the dust content of quiescent galaxies with cosmic time.
In the local Universe, quenched galaxies are overwhelmingly dust poor.
The typical reddening of quiescent galaxies therefore must decline between cosmic noon and the present. 
Building a large sample of well characterized quiescent galaxies at $z\gtrsim1$ is critical to understanding this evolution.
Prospects for constructing such a sample are bright. 
JWST NIRCam covers a wide wavelength range ($0.7-4.8~${\microns} observed-frame) and can densely sample SED shape, with 11 medium- and 9 broad-bands;
as demonstrated in this work, NIRCam photometry successfully limits the dust--age degeneracy, even at $z \gtrsim 2$.
Complementing JWST imaging, 
upcoming multi-fiber spectrographs will efficiently measure galaxy properties and redshifts beyond $z\gtrsim2$---e.g., the Subaru Prime Focus Spectrograph   \citep[PFS,][]{Takada2014,Greene2022} and the VLT Multi-Object Optical and Near-infrared Spectrograph \citep[MOONS,][]{Cirasuolo2014}.
In the near future, these instruments will fill in the picture of galaxy evolution across cosmic time.

\acknowledgments

The BGU lensing group acknowledges support by grant No.~2020750 from the United States-Israel Binational Science Foundation (BSF) and grant No.~2109066 from the United States National Science Foundation (NSF), by the Israel Science Foundation Grant No.~864/23, and by the Ministry of Science \& Technology, Israel.

This research was supported in part by the University of Pittsburgh Center for Research Computing, RRID:SCR\_022735, through the resources provided. Specifically, this work used the H2P/MPI cluster, which is supported by NSF award number OAC-2117681.

DM and RP acknowledge funding from JWST-GO-02561-013 and JWST-GO-04111.035, provided through a grant from the STScI under NASA contract NAS5-03127.

JS acknowledges support by the National Science Foundation Graduate Research Fellowship Program under Grant DGE-2039656. 
Any opinions, findings, and conclusions or recommendations expressed in this material are those of the author(s) and do not necessarily reflect the views of the National Science Foundation.

Support for this work was provided by The Brinson Foundation through a Brinson Prize Fellowship grant.

This work is based in part on observations made with the NASA/ESA/CSA \textit{James Webb Space Telescope}. The data were obtained from the Mikulski Archive for Space Telescopes at the Space Telescope Science Institute, which is operated by the Association of Universities for Research in Astronomy, Inc., under NASA contract NAS 5-03127 for JWST. These observations are associated with JWST Cycle 1 GO program \#2561 and Cycle 3 GO program \#4111. The JWST data presented in this article were obtained from the Mikulski Archive for Space Telescopes (MAST) at the Space Telescope Science Institute. The specific observations analyzed can be accessed via \dataset[DOI]{https://doi.org/10.17909/qgpt-c913}. Support for program JWST-GO-2561 was provided by NASA through a grant from the Space Telescope Science Institute, which is operated by the Associations of Universities for Research in Astronomy, Incorporated, under NASA contract NAS5-26555.

\appendix

\section{Spatially Resolved Analysis}
\label{sec:spatially_resolved_appendix}
\begin{figure*}[t]
\gridline{\fig{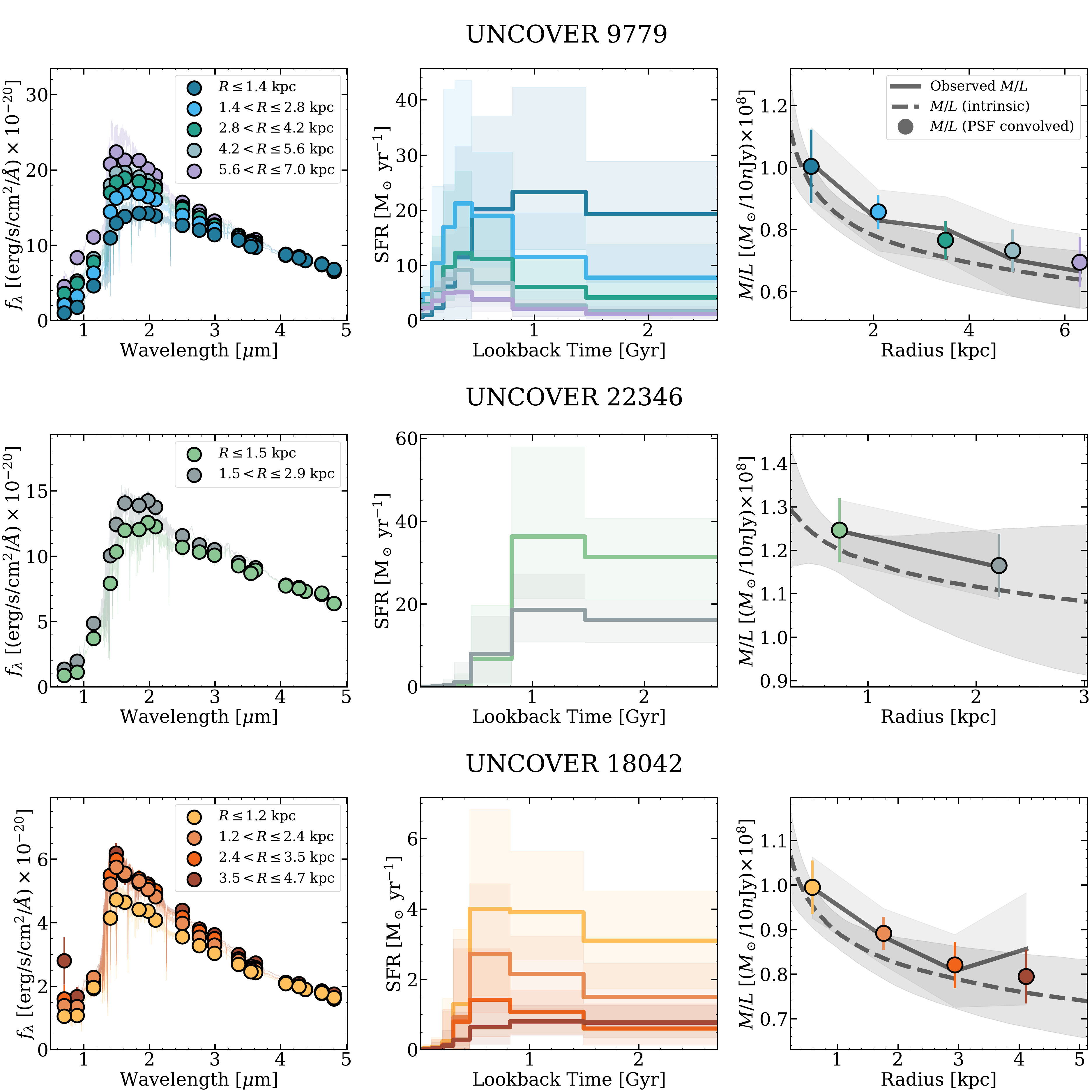}{
\textwidth}{ } }
\caption{
Summary of the annular SED fitting process for the three galaxies not presented in Figure~\ref{fig:methods}. 
Left: the annular SEDs (matched to the central annuli's F444W flux) and their corresponding \texttt{Prospector} fits.
Center: each annuli's star formation history.
Right: the observed mass-to-light $M/L$ ratios (solid line) for each annulus alongside the intrinsic $M/L$ profile (dashed line) and the PSF convolved $M/L$ profile (circles).
The shaded regions and errorbars demarcate $1\sigma$ uncertainties.
} 
\label{fig:methods_appendix}
\end{figure*}

In this work, we perform spatially resolved stellar population synthesis for a sample of  five massive quiescent galaxies.
In Figure~\ref{fig:methods} we diagram our methods for two representative galaxies.
The annular SEDs, star formation histories, and mass-to-light profiles for the remaining three galaxies are presented in Figure~\ref{fig:methods_appendix}.

\section{Comparison Samples}
Our sample of five massive quiescent galaxies at $z \sim 2.5$ are predominately dusty. 
To trace the typical reddening of quenched galaxies as a function of cosmic time, we consider two comparison samples: (i) the \cite{Suess2019b} 3D--HST sample ($0.5 \lesssim z \lesssim 2.5$) and (ii) $z < 0.1$ quiescent galaxies from the GAMA survey. 
The reference samples are limited to galaxies with $\log_{10}[ \frac{M}{M_\odot}] \gtrsim 10$ and rest-frame $UVJ$ colors satisfying the \cite{Whitaker2011} quiescence criteria. 
To ensure internal consistency, both samples are modeled with \texttt{Prospector} (described below).

\subsection{3D--HST}
\label{sec:appendix_3dhst}

For a sample of quiescent galaxies at $z\lesssim2.5$, we consider the \cite{Suess2019b} catalog of $\gtrsim 10^4$ galaxies in 3D--HST.
We select all massive ($\log_{10}[ \frac{M}{M_\odot}] >10$) quiescent \citep[based on  rest-frame $UVJ$ colors,][]{Whitaker2011} galaxies from the \cite{Suess2019b} catalog; the resulting sample spans $0.5 \lesssim z \lesssim 2.5$.
Stellar properties and dust attenuation levels are inferred with \texttt{Prospector} \citep{Leja2019}.
Stellar population models \citep[\texttt{FAST},][]{Kriek2009} are also publicly available from \cite{Straatman2016}.
The evolution of $A_V$ with redshift is qualitatively similar between the \texttt{FAST} and \texttt{Prospector} catalogs. 
For internal consistency, we adopt the \texttt{Prospector} models.

\subsection{GAMA}
\label{sec:appendix_gama}
For a low redshift sample, we consider 150 galaxies from the GAMA survey.
Combining optical spectra with broad photometric coverage (GALEX, SDSS, VISTA, WISE, and Herschel), the GAMA sample offers a panchromatic view of $\gtrsim 10^5$ galaxies at $z<0.65$ \citep{Driver2009,Driver2011}. 
Stellar population fits to the NUV and optical photometry (rest-frame $3000-11000~\mathrm{\AA}$) are publicly available \citep{Taylor2011}.\footnote{\url{https://www.gama-survey.org/dr4/}}
To facilitate comparison with our sample, we remodel a subsample of the GAMA galaxies' SEDs with \texttt{Prospector} (following Section~\ref{sec:annular}).

To construct a low redshift comparison sample, we first select quiescent galaxies.
Using the rest-frame $UVJ$ colors from the GAMA stellar population fits, we apply the \cite{Whitaker2011} selection criteria; to ensure the reported $UVJ$ colors are reasonable, we only select galaxies well fit by \cite{Taylor2011}---i.e., goodness-of-fit statistic~$>0.4$.
Of the quiescent sample, we randomly select 150 galaxies that satisfy: (i) $z<0.1$, (ii) absolute $r$ magnitude~$<-21$ (to approximately select massive galaxies), and (iii) net signal-to-noise ratio~$>80$.
The SEDs of the selected galaxies are then remodeled with \texttt{Prospector}.
We only consider the galaxies' global SEDs (i.e., we did not perform spatially resolved fits).
While  modeling limitations required \cite{Taylor2011} to only fit the rest-frame NUV and optical bands, we use all available photometry. 

\texttt{Prospector} modeling confirms the selected GAMA galaxies are indeed quiescent ($\log_{10}[ \frac{\mathrm{sSFR}}{\mathrm{yr}}] = -13.0_{-1.0}^{+1.0}$) and minimally reddened ($A_V=0.06_{-0.04}^{+0.13}$~mag); the sample's distribution of stellar masses is similar to our five UNCOVER galaxies ($\log_{10}[ \frac{M}{M_\odot}]=10.8_{-0.1}^{+0.3}$).

\bibliography{paper}%

\end{document}